# Ultra-Low Latency for 5G - A Lab Trial

Peng Guan, Xi Zhang, Guangmei Ren, Tingjian Tian
Communication Technology Laboratory
Huawei Technologies Co., Ltd.
Email: {peng.guan; panda.zhang; renguangmei; tiantingjian}@huawei.com

Anass Benjebbour, Yuya Saito, Yoshihisa Kishiyama
Radio Access Network Development Department
NTT DOCOMO, INC.
Email:{ benjebbour; yuuya.saitou.fa; kishiyama}@nttdocomo.com

*Abstract* — **In this paper, we introduce a lab trial that is conducted to study the feasibility of ultra-low latency for 5G. Using a novel frame structure and signaling procedure, a stabilized 1.5 *ms* hybrid automatic repeat request (HARQ) round trip time (RTT) is observed in lab trial as predicted in theory. The achieved round trip time is around 5 times shorter compared with what is supported by LTE-Advanced standard.**

## I. Introduction

With the rapid and enormous success of the 4G LTE networks, study on 5G has been ongoing for several years in both academia and industry. Recently, the standardization process for 5G has also been initiated [1]. As defined by the ITU, three types of services are to be provided by 5G, including enhanced mobile broadband (eMBB), massive machine type communications (mMTC), and ultra-reliable and low-latency communications (URLLC). Similar to the existing MBB service, the eMBB service is to offer an even higher peak data rate (*e.g.*, 20Gbps) for a wide range of deployment scenarios. As for the new services, mMTC is to support an extremely large number of connections (*e.g.*, $10^6$ devices per $km^2$) such as wireless sensors and metering, while URLLC is to provide ultra-low latency (*e.g.*, about 1 *ms* round trip time, RTT) and high reliability data transmission in mission-critical applications. Indubitably, our life can be made even more convenient and enjoyable through these new/upgraded communication services. It can be expected that to efficiently support these new network functionalities, innovative technique components are needed. Studies on 5G concept and candidate technical solutions have been conducted in recent years [2-8]. Moreover, much effort has also been taken to bring these theoretical concepts into practice, which hopefully will serve as cornerstones to enable the three types of services envisaged by ITU.

The emerging new services, such as self-driving cars, industrial control and real-time gaming, are the driving forces behind the requirements on crucial latency and reliability in 5G (e.g., less than 1 *ms* [9]). The hybrid automatic repeat request (HARQ) round trip time (RTT) of LTE-Advanced [10-12] is at the order of ~10 *ms* which is still far away from meeting these requirements. An example is intelligent traffic management system in future smart cities. To guarantee the safety of the self-driving cars, when a braking command is sent, it must be received and acknowledged within a short time duration and proper actions must be taken in time. The latency of LTE-Advanced networks would however lead to an unacceptable displacement from the time the command was initiated to that the command was executed [2]. Another example is real-time gaming. Since the tactile response time for human beings [13] is only several milliseconds, if a ~10 *ms* round-trip latency was implemented, it would lead to a clearly noticeable displacement, which affects the user experience. The example list continues. Thus, the ~1 *ms* latency is an important enabler for various emerging services. And, it brings possibilities for new applications and products, which may revolutionize every aspect of our society such as mobility, traffic, health care, sports, and entertainment.

On the other hand, it is quite a challenging task to ensure such a low latency level within practical communication systems, for the sake of, *.e.g.*, the processing delay at the transmitter and receiver, the propagation delay, the time consumed by possible retransmissions, etc. Recently, latency improvements in further 5G networks have received an increased attention [14-17]. In [14], the authors presented a low latency radio interface design for local area communications by using 180 and 360 KHz OFDM subcarrier spacing with 256 and 128 points FFT size, respectively. In [15], the authors proposed a frame structure based on 312.5 KHz OFDM subcarrier spacing for the ultra-dense small cell networks to reduce the latency. In [16], the authors proposed to reduce TDD air interface latency by using 60 KHz OFDM subcarrier spacing and reducing the frame length to 0.25 *ms*, and it has also been pointed out that a longer battery duration can be expected using such a short radio frame. In [17], the authors provided an overview of the possible access technologies for 5G and some basic trial concepts. In particular, they suggested to use 75 KHz OFDM subcarrier spacing to improve the latency performance. However, all these notable contributions are still highly conceptual and the latency performance in practice remains unclear.

Taking all these into consideration, the major contribution of this paper is a lab trial with the state-of-the-art hardware to validate the feasibility of the proposed concept. In this trial, a new frame structure and the associated signaling procedure are proposed to reduce the transmission latency, in terms of the HARQ RTT, to a level around 1 *ms*. To the best of the authors' knowledge, this is the first reported practical trial for low latency applications in 5G. By the proposed frame structure, the trial results confirm around 5 times shorter RTT latency compared to the existing LTE-Advanced standard.

The rest of this paper is organized as follows. Section II focuses on the new frame structure and the associated HARQ procedure used in the lab trial. Section III introduces the test bed hardware used for the lab trial. Section IV describes the details and summarizes the results of the lab trial conducted. Finally, Section V concludes this paper with plans for future work.

## II. PROPOSED FRAME STRUCTURE AND HARQ PROCEDURE

In this section, we first revisit the framework structure and latency performance in terms of HARQ RTT in LTE-Advanced. Then the new frame structure used in the 5G lab trial is introduced and discussed. With this new frame structure, a much shorter of about 1.5 *ms* HARQ RTT is expected for TDD downlink. In this lab trial, TDD mode and OFDM modulation scheme are assumed.

### A. Review of LTE-Advanced Frame Structure

In 3GPP LTE-Advanced specifications [10], OFDM is the modulation scheme used in the downlink. The subcarrier spacing $\Delta f$ is 15 KHz, the OFDM symbol duration $T_{\mathrm{OFDM}}$ is $1/\Delta f$ = 66.67 *us*, the FFT size $N_{\mathrm{FFT}}$ is 2048, the sampling rate $f_s$ is $\Delta f \times N_{\mathrm{FFT}}$ = 30.72 MHz, the sample interval $T_s$ is $1 / f_s$. The LTE frame duration $T_f$ is 10 *ms*. Each frame is divided into 10 subframes, each of $T_{sf}$ = 1 *ms* duration, which is further divided into two 0.5 *ms* slots in the time domain. Meanwhile, OFDMA is the multiplexing scheme used in the downlink, and UEs are allocated with certain numbers of physical resource blocks (PRBs) by the scheduler at the BS side. A PRB spans 0.5 *ms* in the time domain, *i.e.*, 6 or 7 ODFM symbols (depending on whether the normal or extended CP was used) and 180 KHz in the frequency domain, *i.e.*, 12 consecutive OFDM subcarriers.

As an indicator of the latency performance, HARQ RTT Timer specifies the minimum number of subframes before a downlink HARQ retransmission is expected by the UE [12]. It is usually assumed that either the BS or the UE needs about 3 *ms* to process the data received and prepare the Acknowledgement (ACK) or Negative-ACK (NACK) message and 1 *ms* for one-way data or HARQ transmission. Thus, for FDD mode, the HARQ RTT Timer is set to 8 subframes. For TDD mode, the HARQ RTT Timer is set to $k + 4$ subframes, where $k$ (usually 6 or 7) is the interval between the first downlink transmission and the transmission of the associated HARQ feedback. Figure 1 and Figure 2 provide illustrations of the 8 *ms* HARQ RTT in the FDD downlink and the 11 *ms* HARQ RTT in the TDD downlink, respectively.

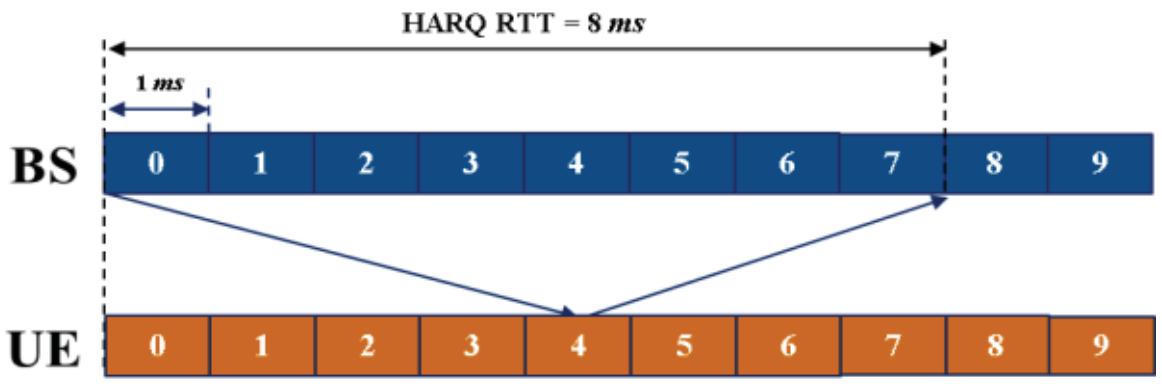

Figure 1. HARQ RTT for FDD downlink in LTE-A

### B. Proposed Frame Structure

As can be seen in the last subsection, the latency in LTE-Advanced standard is largely limited by its frame structure and the minimum number of subframes before a downlink HARQ retransmission is expected. To achieve the goal of ~1 *ms* HARQ RTT in TDD-based networks, a new frame structure is needed. Under such a time budget, a natural solution is to reduce the subframe duration. This can be obtained by keeping the OFDM symbol duration and reducing the number of OFDM symbols in each subframe, or, keeping the number of OFDM symbols in each subframe and reducing the OFDM symbol duration. In this work, we take the second approach. That is, the subcarrier spacing is enlarged to shorten the OFDM symbol duration and the number of OFDM symbols in each subframe is kept unchanged in the new frame structure for TDD downlink. More specifically, the subcarrier spacing $\Delta f$ is changed to 30 KHz, the corresponding OFDM symbol duration $T_{\text{OFDM}}$ becomes 33.33 *us*, and the FFT size $N_{\text{FFT}}$ is now 1024, whilst the sampling rate $f_s$ is kept same as 30.72 MHz which is identical to that in LTE-Advanced. OFDMA is used as the downlink multiplexing scheme. The frame duration $T_s$ is still 10 *ms*. It can be divided into 40 subframes. Each subframe duration is $T_{\text{sf}}$ = 0.25 *ms*, and it contains 7 OFDM symbols in the time domain, as indicated in Figure 3.

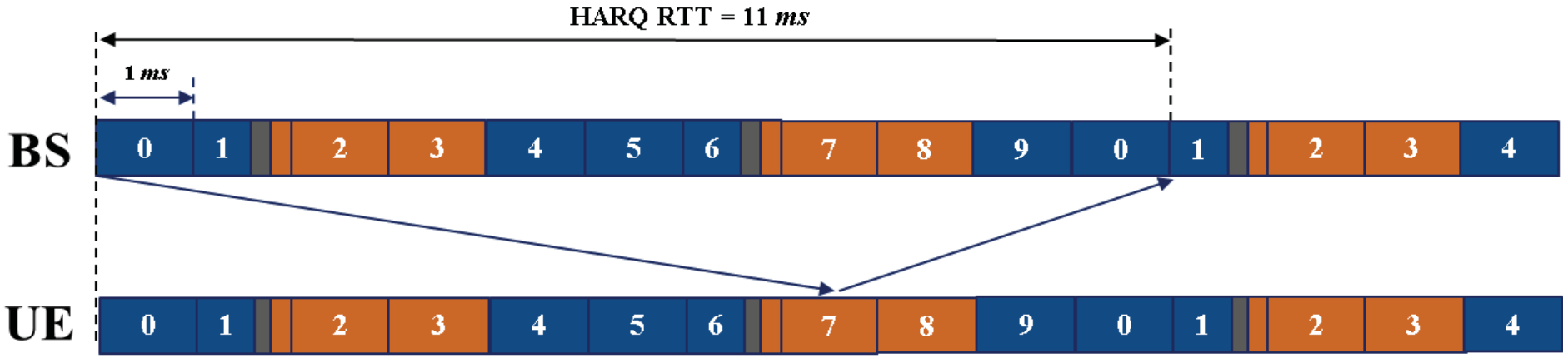

Figure 2. HARQ RTT for TDD downlink in LTE-Advanced

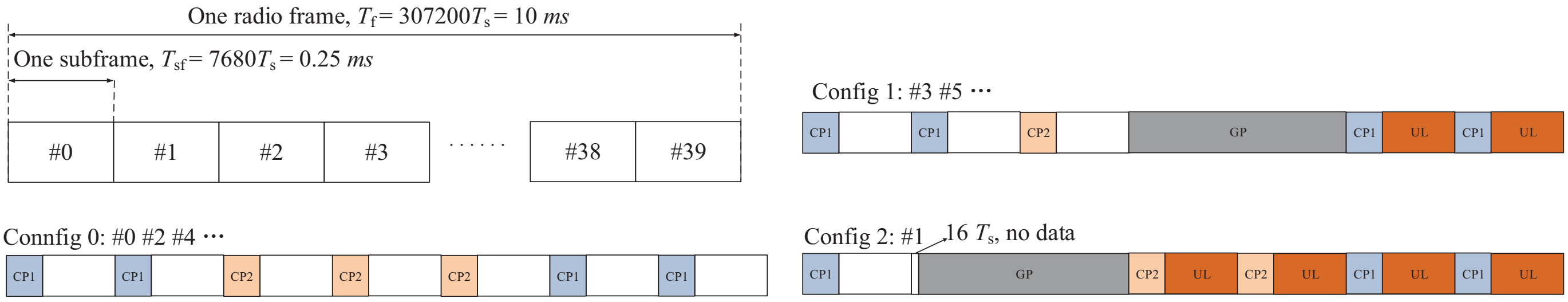

Figure 3. Proposed frame structure

Two types of CPs are employed with the durations $T_{\text{cp1}} = 5/64 \times N_{\text{IFFT}} \times T_s \approx 2.604$ *us* and $T_{\text{cp2}} = 4/64 \times N_{\text{IFFT}} \times T_s \approx 2.083$ *us*, respectively. The proposed frame structure can be configured in a flexible manner. Different modes can be distinguished by the configuration index, as shown in Table 1. Each configuration corresponds to a different ratio among the numbers of OFDM symbols used for downlink (DL), uplink (UL) and guard period (GP) between downlink and uplink switching. Among them, configuration 1 can be considered as a self-contained frame structure, which involves both DL and UL transmission, two OFDM symbols are used for UL transmission (e.g., ACK/NACK feedback), and can also be configured to other values. It is also worth mentioning that Configuration #2 is specially designed for uplink access.

Table 1 Subframe Configuration

| *Configuration Index* | *DL:GP:UL* | *Notes* |
|---|---|---|
| 0 | 7:0:0 | Pure DL |
| 1 | 3:2:2 | Two UL symbols |
| 2 | 1:2:4 | For UL access |

### C. 1.5 ms HARQ RTT

In addition to a shorter subframe duration, the tight RTT budget, *e.g.*, ~1 *ms*, also asks for a substantial reduction on the processing delay at both the BS and UE side. For the services of interests like industrial control and self-driving cars, the data packet would be made small to reduce the processing delay and achieve a high reliability.

Based on the new frame structure proposed in the last subsection and assuming fast processing and transmission at the BS and UE, the HARQ RTT Timer for TDD downlink is set to 6 subframes, *i.e.*, 1.5 *ms*, as illustrated in Figure 4. More specifically, the physical layer procedure is detailed in Figure 5 and the 1.5 *ms* HARQ RTT can be broken into:

- Downlink transmission time: ~0.25 *ms*
- UE processing time (max.): ~0.7 *ms*
- ACK/NACK transmission time: ~0.07 *ms*
- BS processing time (max): ~0.5 *ms*

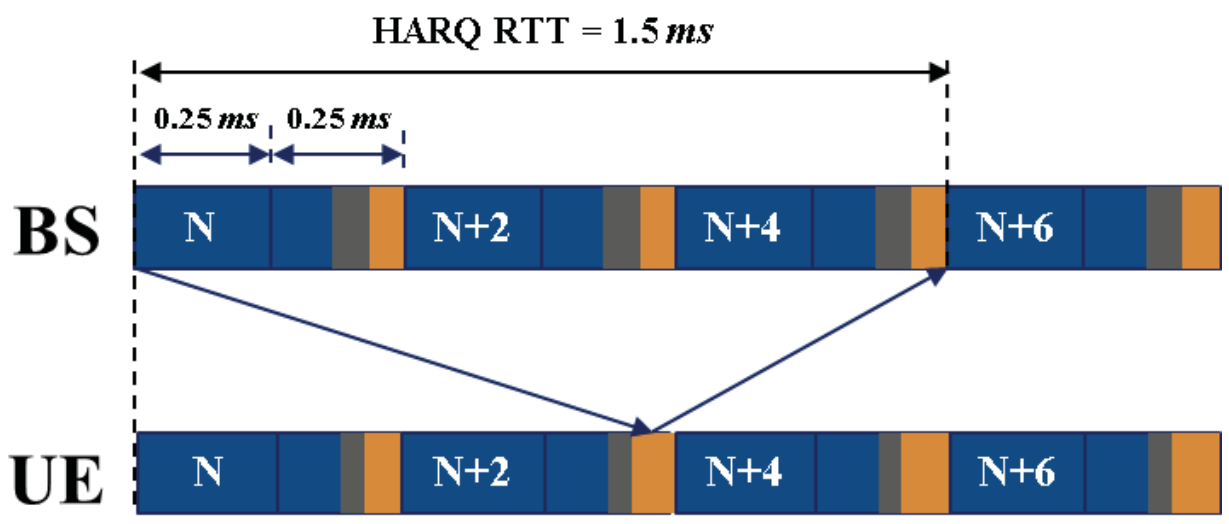

Figure 4. Proposed 1.5 *ms* HARQ RTT for TDD downlink

## III. Test Bed Architecture

In this section, the test bed hardware used for lab trial is described, which consists of a BS and a UE. As illustrated in the last section, the challenges for achieving a low latency mostly come from the transmission delay and the processing delay. While the transmission delay can be reduced by revising the frame structure, e.g., shortening the subframe duration, the processing delay can be made small by leveraging the recent advancements in computing technologies. We anticipate that a properly designed frame structure and the associated signaling procedure, and a well-chosen packet size as well as the constantly improved processing capacity will the key enablers for bringing 5G low latency applications into practice.

The BS unit in the test bed used for the lab trial, as depicted in Figure 6, has three major units: a baseband unit, a digital IF unit and an analog IF & RF unit. The baseband unit employs a high-performance computing server, and it is responsible for baseband signal processing, scheduling, channel coding, MIMO precoding, and OFDM symbol generation, etc. The digital IF unit, implemented on high-performance FPGA chipsets, is responsible for digital up conversion (DUC) / digital down conversion (DDC), filtering, crest factor reduction (CFR) and digital pre-distortion (DPD), etc. The analog IF & RF unit adopts a high-performance frequency mixer and a high-efficiency wideband power amplifier (PA). It supports up to 4.6 GHz carrier frequency and the maximum output power is 28 dBm per channel. Optical fibers are used for connection between the baseband unit and the digital IF unit, as well as the connection between the digital IF unit and the analog IF & RF unit.

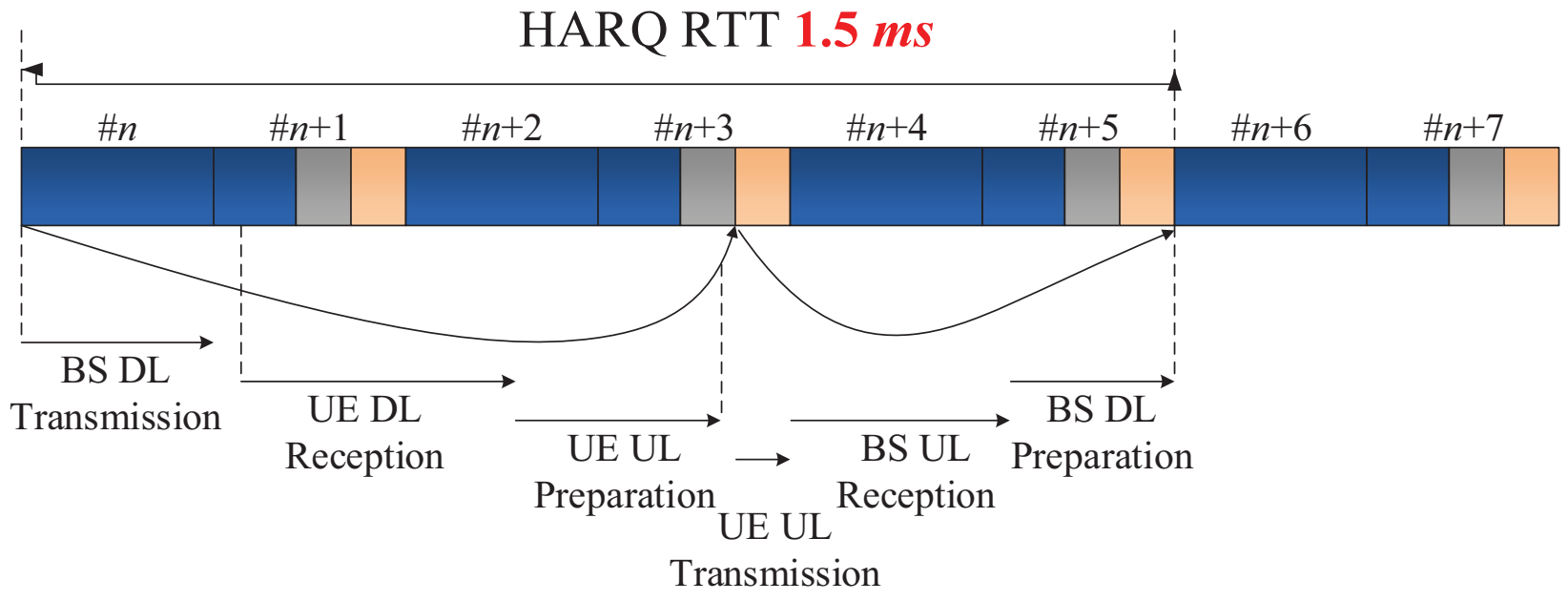

Figure 5. Physical layer procedure of 1.5 *ms* HARQ RTT for TDD downlink

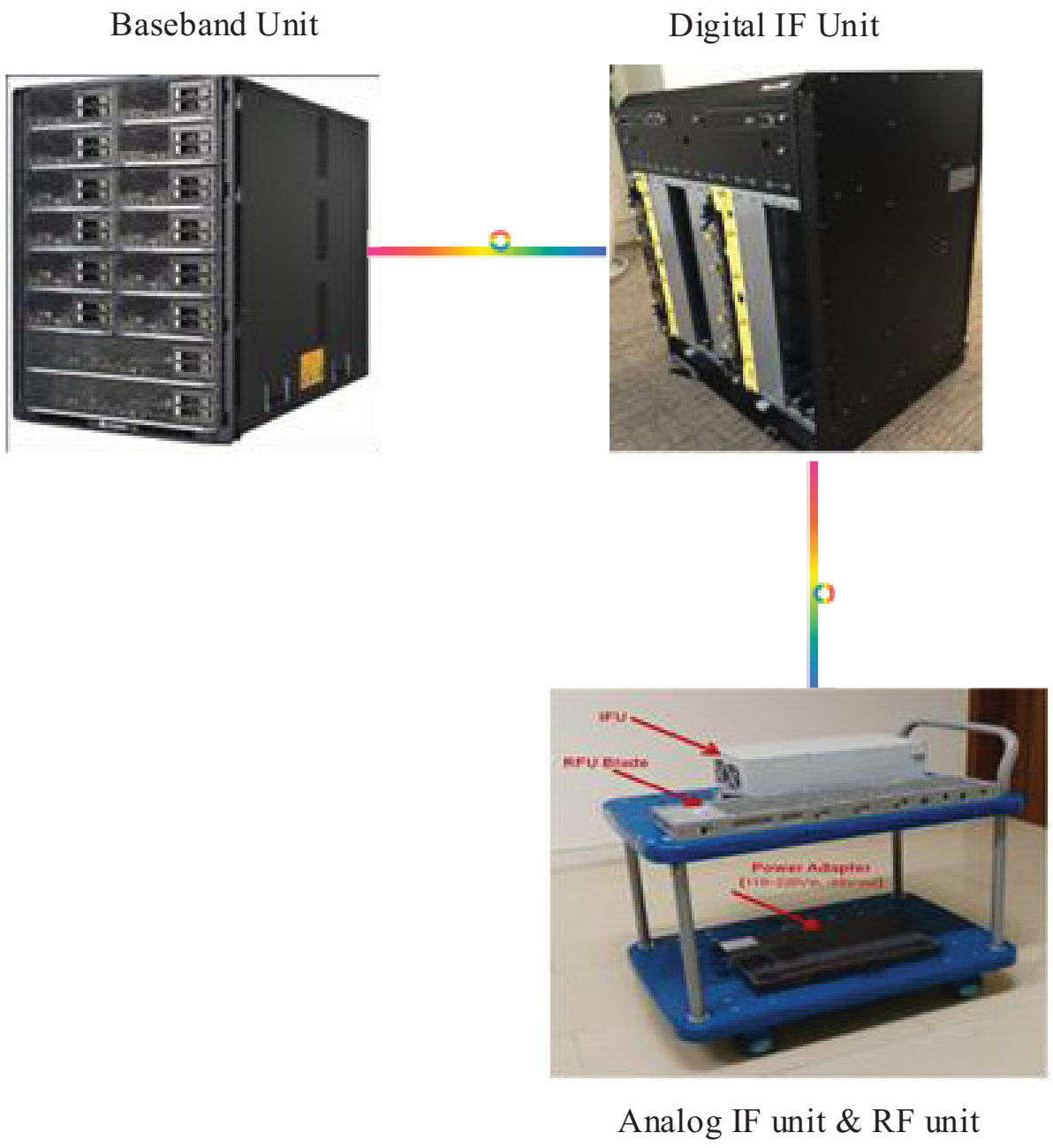

Figure 6. BS hardware

The UE hardware, as depicted in Figure 7, integrates a RF unit and a baseband unit with a high-performance computing server to process the functionalities including downlink MIMO detection, channel decoding and uplink ACK/NACK feedback transmission, etc.

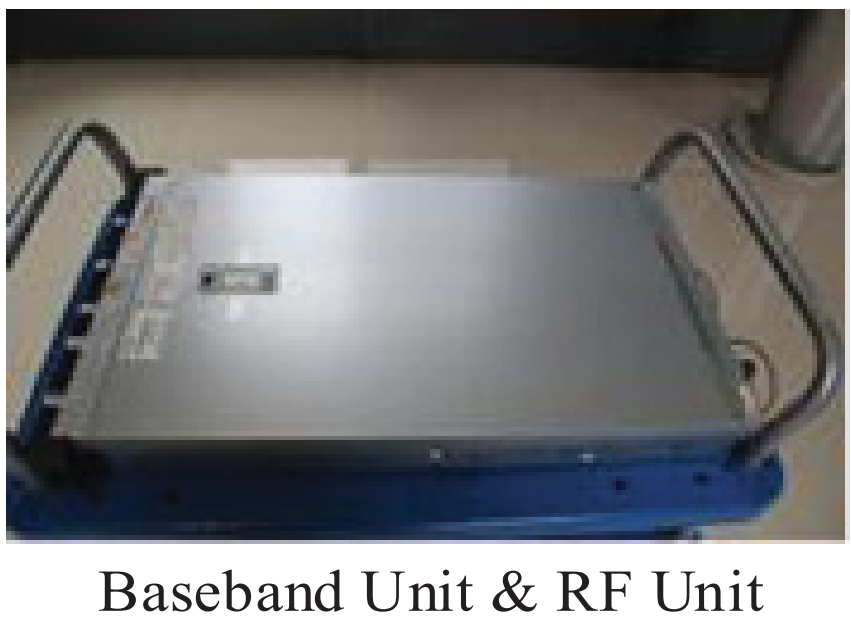

Figure 7. UE hardware

## IV. Lab Trial Setup and Results

In this section, we present the details of the lab trial conducted to validate of the feasibility and reliability of the low latency concept based on the proposed frame structure and the HARQ round trip procedure, using the test bed introduced in Section III. Firstly, we illustrate the measurement setup and the exact parameterization used in lab trial. And then, the results are summarized and analyzed, which, as expected, confirms the achievability of ~1.5 *ms* HARQ RTT for TDD downlink.

### *A. Measurement and Parameterization*

The lab trial is carried out in an indoor environment (an anechoic chamber located in the NTT DOCOMO R&D center in Japan) as shown in Figure 8. With the test bed introduced in Section III, the BS and the UE are both equipped with 2 transmit antennas and 2 receive antennas to apply the MIMO transmission with space-frequency block coding (SFBC), while the modulation and coding scheme (MCS) is kept fixed for all transmissions. The lab trial also assumes a short transmission range between the BS and the UE and the variation in the received signal-to-noise-ratio (SNR) is realized through the adjustment of the output power at the BS analog IF unit.

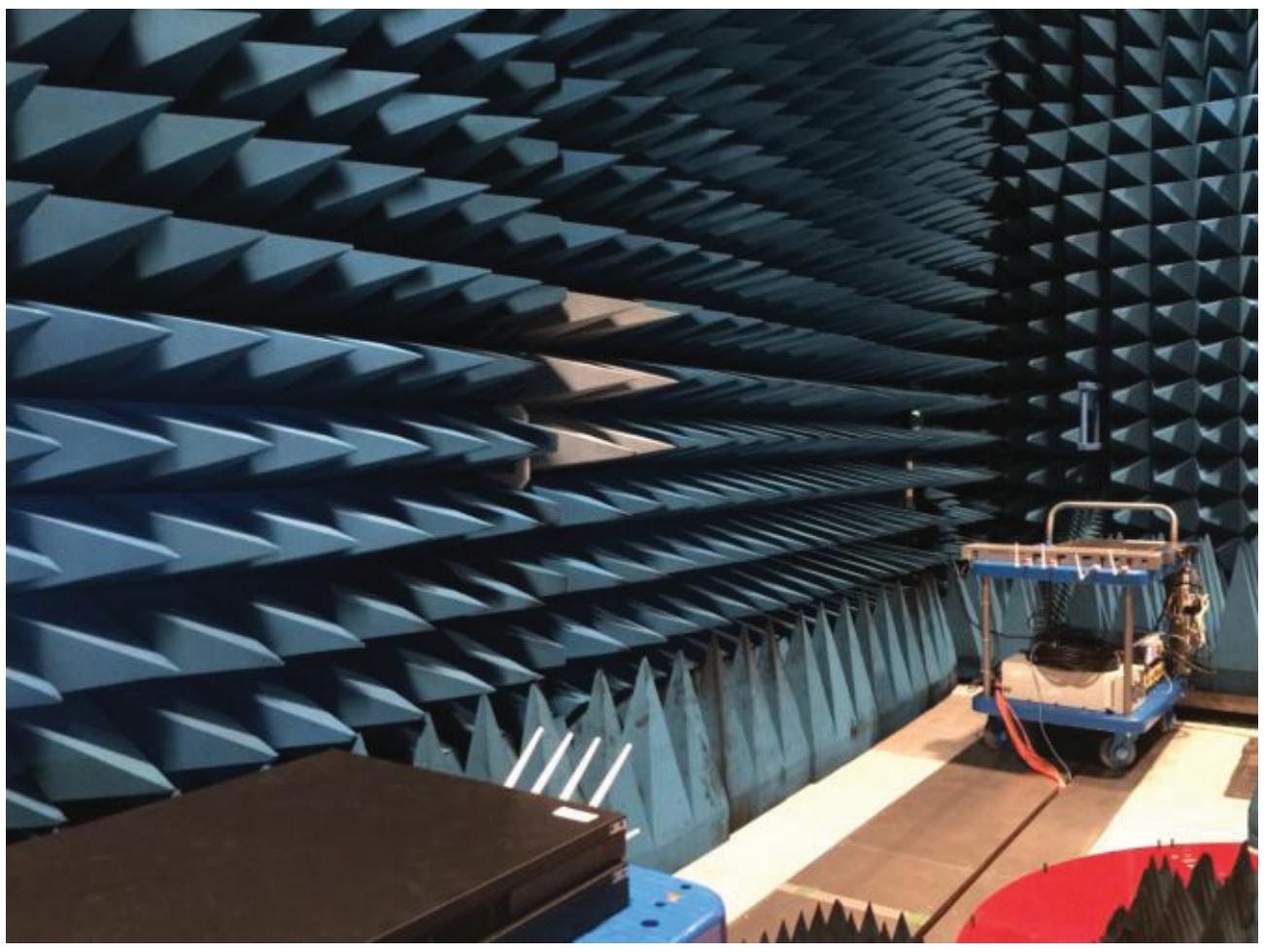

Figure 8. Lab Trial Environment

More specifically, the parameters used in this lab trial are listed in Table 2. It is worth emphasizing the major differences from the conventional LTE-Advanced setup including a higher carrier frequency ($f_c$ = 4.6 GHz) and a shorter OFDM symbol duration ($T_{OFDM}$ = 33.33 *us*) corresponding to a larger subcarrier spacing ($\Delta f$ = 30 KHz). During the trial, 0.72 MHz out of the total 20 MHz carrier bandwidth is dedicated to the delay-sensitive and mission-critical applications by the scheduler at the BS side. The downlink packet formation, *e.g.*, the relationship between the MCS and the transport block size (TBS), are determined in consistence with the LTE specifications. However, the number of resource elements (REs) embedded in each PRB is different from that in the LTE-Advanced specifications.

Table 2 Parameters used in Lab Trial

| ***Parameter*** | ***Value*** |
|---|---|
| Carrier frequency | 4.6 GHz |
| Carrier bandwidth | 20 MHz |
| Bandwidth for low-latency service | 0.72 MHz |
| Subcarrier spacing | 30 KHz |
| OFDM symbol duration | 33.33 *us* |
| Number of UEs | 1 |
| Number of layers | 1 |
| BS antennas | 2T2R |
| UE antennas | 2T2R |
| Downlink MIMO mode | SFBC |
| MCS | Fixed |

### *B. Results Analysis*

The measurement in the lab trial focuses on the successful decoding rate and the ACK/NACK loss ratio when different fixed MCSs are used. The results are summarized in the Table 3 and Table 4 for high SNR (25 – 26 dB) and low SNR (6 – 8 dB) regions, respectively.

It can be observed that, the ACK/NACK loss ratio is always 0%, which means that in each HARQ round trip for TDD downlink, the BS has successfully received the ACK/NACK messages and has effectively sent the next queuing data to the UE in the specified time limit. Thus it is safe to say that the expected goal on latency, *e.g.*, 1.5 *ms* HARQ RTT for TDD downlink, has been achieved, regardless of the SNR and MCS configurations.

Table 3
Decoding rate and ACK/NACK loss ratio in the high SNR region with different MCSs

| ***MCS*** | ***Packet size [bits]*** | ***Decoding rate [%]*** | ***A/N Loss [%]*** |
|---|---|---|---|
| 8 (QPSK) | 120 | 100 | 0 |
| 14 (16 QAM) | 224 | 100 | 0 |
| 16 (16 QAM) | 328 | 100 | 0 |
| 20 (64 QAM) | 376 | 100 | 0 |
| 26 (64 QAM) | 584 | 99.98 | 0 |
| 28 (64 QAM) | 712 | 99.93 | 0 |

Table 4
Decoding rate and ACK/NACK loss ratio in the low SNR region with different MCSs

| ***MCS*** | ***Packet size [bits]*** | ***Decoding rate [%]*** | ***A/N Loss [%]*** |
|---|---|---|---|
| 8 (QPSK) | 120 | 96.69 | 0 |
| 14 (16 QAM) | 224 | 91.79 | 0 |
| 16 (16 QAM) | 328 | 84.71 | 0 |
| 20 (64 QAM) | 376 | 83.04 | 0 |
| 26 (64 QAM) | 584 | 10.65 | 0 |
| 28 (64 QAM) | 712 | 0.34 | 0 |

Also, the decoding rate coincides with our intuitions. To be specific, in the high SNR region, the UE can almost always successfully decode the transmitted messages even with high MCSs. In the low SNR region, the decoding rate reduces with increasing the MCS index, corresponding to high-order modulation and higher coding rate.

## V. Conclusions and Outlook

In this paper, we reported a stably achievable 1.5 ms HARQ RTT for TDD downlink in a lab trial with the state-of-the-art test equipment using a novel frame structure and the associated signaling procedure. Compared with the existing LTE-Advanced standard, around 5 times better latency performance was obtained and it makes a good start for bringing the low latency feature of 5G networks into practice.

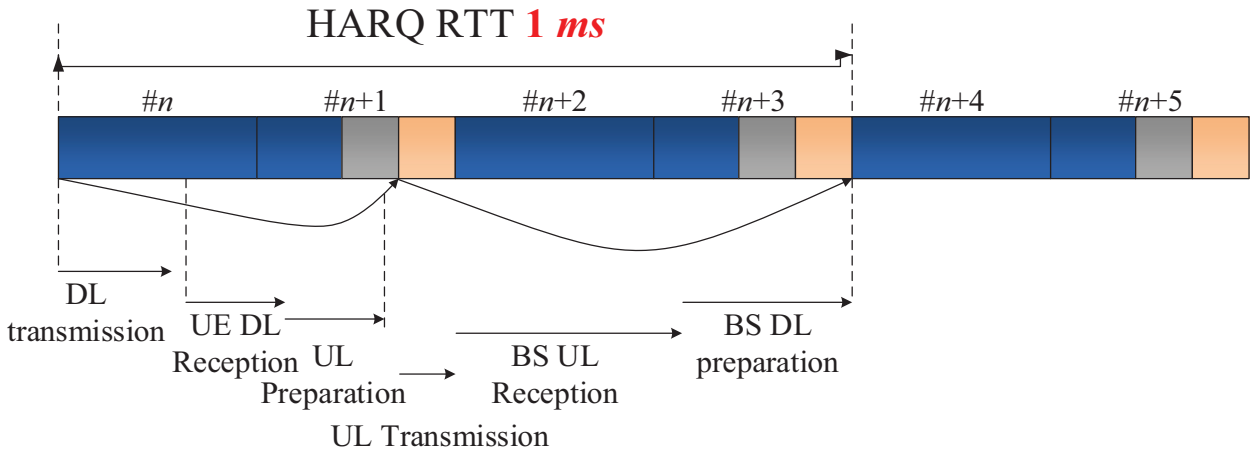

Figure 9. Proposed 1 *ms* HARQ RTT for TDD downlink

As for future work, 1.5 *ms* HARQ RTT for TDD uplink is scheduled to be tested. The even lower latency, *e.g.*, 1 *ms*, is planned to be verified too, possibly using a signaling procedure as illustrated in Figure 9, *i.e.*, with an even faster ACK/NACK feedback. In a longer scope, the feasibility studies, lab trials and field trials are expected to continue for more practical scenarios.